# Recent Advances in Electron and Positron Sources[*]

J.E. Clendenin

*Stanford Linear Accelerator Center*
*Stanford, CA 94309*

**Abstract**

Recent advances in electron and positron sources have resulted in new capabilities driven in most cases by the increasing demands of advanced accelerating systems. Electron sources for brighter beams and for high average-current beams are described. The status and remaining challenges for polarized electron beams are also discussed. For positron sources, recent activity in the development of polarized positron beams for future colliders is reviewed. Finally, a new proposal for combining laser cooling with beam polarization is presented.

*Invited talk presented at*
*The 9th Workshop on Advanced Accelerator Concepts*
*Sante Fe, NM*
*10-16 June 2000*

[*]Work supported by Department of Energy contract DE-AC03-76SF00515.

# Recent Advances in Electron and Positron Sources*

J.E. Clendenin

*Stanford Linear Accelerator Center*
*Stanford, CA 94309*

**Abstract.** Recent advances in electron and positron sources have resulted in new capabilities driven in most cases by the increasing demands of advanced accelerating systems. Electron sources for brighter beams and for high average-current beams are described. The status and remaining challenges for polarized electron beams are also discussed. For positron sources, recent activity in the development of polarized positron beams for future colliders is reviewed. Finally, a new proposal for combining laser cooling with beam polarization is presented.

## ELECTRON SOURCES

Throughout the long history of electron sources, the standard and virtually universal configuration of the dc-biased gun has been a coaxial design with a ceramic high-voltage insulating tube providing the outside vacuum envelope and inside, supporting the cathode, a metal tube at atmospheric pressure as illustrated in Fig. 1(a). This design works well in practice, but the insulator is large, and the outside of the ceramic is subject to contamination from the atmosphere, which can lead to excessive leakage current. With the advent of polarized electron sources, various vacuum components associated with installing cathodes under vacuum are typically attached to the high-voltage flange, leading to an awkwardly large high-voltage assemblage [1]. These problems are solved for photocathode dc-biased guns by inverting the structure to have a smaller-diameter ceramic tube inside a larger metal vacuum chamber as shown in Fig. 1(b). The earliest such designs were developed in the early 1990s at SLAC [2] and independently at Novosibirsk [3]. The SLAC design was built and successfully tested at high voltage but has not yet been used to produce electrons. Variations on the basic idea include the double insulator design, Fig. 1(c), used at Amsterdam [4] and under development at Nagoya [5], and variations on the high-voltage connection, Fig. 1(d), as used at Mainz [6] and Bonn [7]. For a pulsed high-voltage gun, the Fig. 1(b) design looks most promising.

The past decade has seen the rapid development of photocathode rf guns [8]. These guns are especially well-suited as high-brightness sources. Since the emittance requirements of future colliders seems beyond the reach of any rf photoinjector design, the need for high-brightness rf guns comes mostly from free electron laser (FEL) developments. As an example, the Linac Coherent Light Source (LCLS) requires a 1-nC, 100-A beam from the photoinjector with a normalized rms transverse emittance, $\varepsilon_{n,rms}$, of 1 µm [9]. The emittance growth in an rf photoinjector is mostly correlated. Special techniques have been developed to reverse this growth and then lock-in the resulting emittance minimum just as the beam becomes relativistic. Experimentally the lowest emittance for an LCLS type beam is $\varepsilon_{n,rms}$~2 µm [10,11]. However, these measurements were done with a spatially uniform but temporally Gaussian charge distribution. Simulations using PARMELA, a multi-particle tracking code, indicate that if the temporal distribution is also uniform, $\varepsilon_{n,rms}$~1 µm should be achievable. Progress in exploring the relevant parameter space has been facilitated by the recent development of a semi-analytic code, HOMDYN [12]. Using newly discovered matching conditions [13], simulations now predict an emittance for an LCLS type beam of close to $\varepsilon_{n,rms}$~0.5 µm (thermal effects included) at 150 MeV [14]. See Fig. 2.

*Work supported by Department of Energy contract DE-AC03-76SF00515.

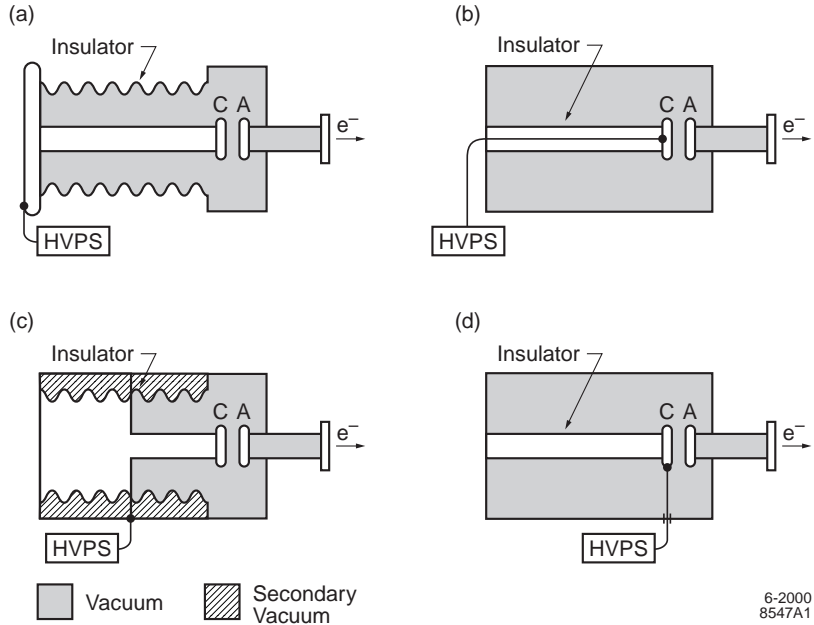

**FIGURE 1**. The inverted structure (IS) design and variations: (a) the conventional non-inverted design; (b) the original IS design; (c) the double insulator design; and (d) variation on connecting the high-voltage power supply (HVPS). For the IS design, the cathode (C) and anode (A) electrodes are permanent, while the photocathode (not separately shown) is removable.

The possibility of achieving even higher brightness using a pulsed diode-structure photocathode gun with GV/m level fields was first reported in 1996 [15]. The high fields are achieved by using a pulsed voltage on the order of 1 MV across a gap of 0.5 to 1 mm. Very short voltage pulses on the order of nanoseconds are used to minimize breakdown and field emission. Ideally the laser pulse should be shorter than the voltage pulse. For an LCLS-type pulse, simulations predict an emittance of 0.4 µm measured 3.25 cm from the cathode [16]. Since space charge effects are still significant at 1 MeV, a design for matching this beam into an accelerating structure is needed to evaluate the final emittance at high energy.

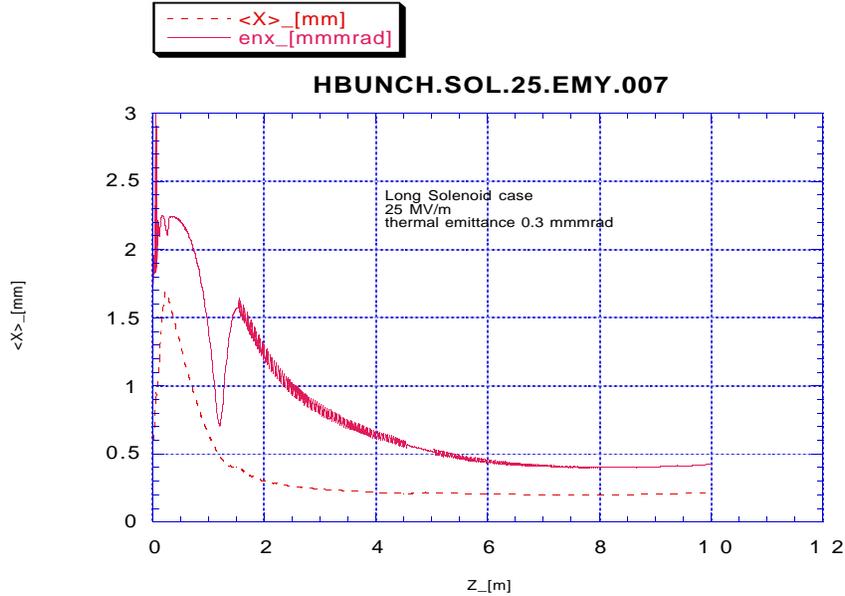

**FIGURE 2.** Transverse normalized rms emittance and beam size, computed using HOMDYN, for the LCLS photoinjector, which consists of a 1.6-cell S-band gun with cathode at z=0 and two 3-m TW sections beginning at z~1.5 m [14]. The peak field in the gun is 130 MV/m, while the sections, with a weak solenoid around the first, are operated at 25 MeV/m. Thermal emittance for a Cu cathode is included.

Photocathode guns using III-V semiconductor cathodes are now universally used as sources of polarized electron beams for accelerators. The successful operation of such beams at SLAC, Mainz, JLAB and elsewhere has demonstrated operating parameters (not all achieved at the same time) well matched to accelerator requirements. Some of these parameters are shown in Table 1.

There remain at least 3 challenges for future polarized electron sources: overcoming the cathode charge limit; increasing the polarization, $P$; and increasing the average current. 1) The maximum current density that can be extracted is limited by a surface barrier that dynamically grows when charge is temporarily trapped at the surface faster than it can recombine with holes. For a pulse train, such as required by most future collider designs, each pulse is influenced by the decaying surface barrier generated by the previous pulses. New cathode structures plus differential doping may solve this problem [18]. 2) Higher polarization will improve the effective luminosity of any high-energy experiment which depends on polarized electrons. In addition, for a future collider, $P>95\%$ may be the only reasonable route to certain new physics. Most of the polarization loss in the cathode bulk can probably be eliminated, but losses in the band bending region may be unavoidable, limiting the maximum polarization to $P\sim90\%$. 3) Finally, the high average currents required by cw accelerators and some types of FELs result in a rapid loss of quantum efficiency (QE) due to ion bombardment at the cathode. Improving the vacuum near the cathode will minimize this effect.

**TABLE 1. Operating Parameters Achieved for Polarized Electron Sources.**

| Parameter | Value | Where Achieved |
|---|---|---|
| Current Density, $J$ | 10 A cm$^{-2}$ | SLAC |
| Average Current, $I_{AVG}$ | 5 mA | JLAB [17], GaAs, unpolarized |
| Polarization, $P$ | 80% | SLAC |
| Cathode 1/e Lifetime, $\tau$ | >1000 h | SLAC |
| Operating time per cathode | >5000 h | SLAC |

Field emitter arrays, ferroelectrics, and secondary electron emitters have the potential to overcome the limitations found with photocathodes for producing high average currents [19]. At the present time, the latter is the most promising. It consists of an rf cavity equipped with a secondary emission surface at one end and a secondary emission grid at the beam exit. Startup electrons multiply rapidly during each rf cycle while simultaneously bunching. Steady state conditions are achieved within a few cycles for a pulse train of fixed charge and pulse length. While the pulse spacing is fixed by the rf frequency, the pulses are automatically synchronous with the rf. Proof of principle testing has been carried out at low charge, but simulations show that the charge per bunch can be up to 500 nC [20]. There appears to be some possibility of modulating the charge using a separate grid.

## POSITRON SOURCES

Conventional positron sources for accelerators use a high-energy electron beam impinging on a high-Z material such as W to generate ~100 MeV γs by bremsstrahlung. The γs in turn create electron-positron pairs in the same material. Positrons exiting the target in the 2-20 MeV regime are confined by a magnetic field while being inserted into an rf accelerating field, bunched and accelerated to relativistic energies for transport to the main linac. Because the initial positron beam emittance is large, more damping is required than for an equivalent intensity electron beam. The NLC positron source design is essentially a scaled version of the SLC source.

It is highly desirable that the positron beam for a future collider be polarized [21]. For many types of experiments, the polarization of the two colliding beams combine to create in effect a single higher polarization. Thus a highly polarized electron beam colliding with a modestly polarized positron beam may be equivalent to the desired single beam polarization $P>95\%$. Another class of experiments is possible only with both beams polarized. At least 3 methods of producing polarized positron beams have been suggested for colliders: helical undulator; Compton scattering; and polarized electrons.

Circularly polarized γs in the required energy range and of sufficient intensity can be produced by passing a 150 GeV electron beam through a 150-m helical undulator [22]. The γs are directed to a thin conversion target placed downstream after the electron primary beam is bent away. This is the design chosen by TESLA [23] but to date rejected by NLC out of concern that the post-interaction beam is too disrupted, and that alternatively having the undulator in the main linac beamline will impose too great an operational restriction on the linac. Recently an interesting proposal has been made to operate the first part of the linac at double the normal rf repetition rate in order to accelerate positron-production electrons along with the main beam, then at the 150-point deflecting the positron-production electrons into a separate beamline having the undulator and target [24].

In the second method, circularly-polarized high-energy γs are produced by Compton backscattering of circularly-polarized photons by unpolarized high-energy electrons. Again the γs are directed to a thin conversion target. Such a scheme was first proposed for the JLC by Okugi et al. in 1996 [25]. Eighty-five $CO_2$ lasers, each producing 10 J per pulse (150 Hz) at the fundamental (10.6 μm) are required, one laser for each micropulse in the JLC pulse train. The cross section for Compton scattering is optimized by choosing a 6.7 GeV electron beam. A similar scheme has been proposed by Frisch (1997) that utilizes an Nd:glass laser (1.05 μm) and 1.7 GeV electrons [26]. In the latter case, in order to reduce the laser energy requirements, a resonant cavity is introduced to recycle the optical power, allowing the same optical pulse to interact with many electron bunches. The mirrors for the optical cavity are problematic because of the high energy in each laser pulse. A recent experiment at the Accelerator Test Facility (ATF) at KEK demonstrated that the production of positrons from a thin conversion target for which the γs were produced by scattering a 200 mJ Nd:YAG laser beam from the 1.26 GeV ATF electron beam was as expected [27].

The third method is a modification of the conventional scheme. If the incident electron beam is highly polarized, then both the high-energy end of the γ spectrum and of the resulting positrons will be polarized [28]. The problem here is that the yield is estimated to be 3 orders of magnitude below that required for colliders. One can imagine a number of ways to increase the total yield, including increasing the charge per pulse in the production beam, filling more rf buckets, using multiple sources, etc., so that in principle the yield might be forced to be adequate, but the practical aspects are daunting.

## SIMULTANEOUS LASER DAMPING/POLARIZING

Potylitsin has recently proposed [29] that direct polarization of an unpolarized positron beam by Compton scattering [30] may be a more efficient source of polarized positrons than the methods discussed above.[1] For this case, the energy and polarization of the polarized positrons after N collisions with identical circularly polarized photons are (here only $\hbar = m_e = c = 1$) [29]:

$$\gamma_{(N)} = \frac{\gamma_0}{1+2\mu} \text{ and } \xi_{(N)} = \frac{\mu}{1+\mu}, \qquad (1)$$

where $\mu = \gamma_o \omega_o N = \frac{4}{3} \frac{A}{m_0 c^2} \gamma_o \left(\frac{r_e}{\sigma_{ph}}\right)^2$ and $\sigma_{ph}^2 = \frac{\lambda_o l_e}{8\pi}$. Here $\gamma_0$ and $\omega_0$ are the initial positron and photon energies, $A$ is the laser flash energy, $r_e$ is the classical electron radius, and $\lambda_o$ and $l_e$ are the laser wavelength and positron bunch length respectively. For example, a 2 GeV positron bunch in a single interaction with a 25 J laser pulse ($\lambda_o$=1 μm) would be expected to result in a polarization of ~60% if $l_e$ can be reduced to 0.2 mm [31]. For a collider such as the NLC with 95 microbunches per pulse train and 120 Hz repetition rate, an average laser power of 0.3 MW is required! However, since the positron beam must in any case be damped, let us review the requirements for laser damping.

---
[1] The process works equally well for electrons.

A powerful laser can be used to damp an electron or positron beam by Compton scattering [32]. The requirement for a significant reduction in the initial transverse normalized emittance, $\varepsilon_{no}$, is that the electrons should lose a similar fraction of their initial energy, $E_o$, as a result of the Compton interaction:

$$\frac{\varepsilon_{no}}{\varepsilon_n} \cong \frac{E_o}{E} = 1 + \frac{64\pi^2 r_e^2 \gamma_o}{3 m_o c^2 \lambda l_e} A, \qquad (2)$$

where $A[J] = \frac{25\lambda[\mu m] l_e[mm]}{E_o[GeV]} \left( \frac{E_o}{E} - 1 \right)$. For $E_o$=2 GeV, $A$=10 J (at $\lambda$=1 μm) is required in a single pass to reduce the transverse emittance by a factor of 10 (again assuming $l_e$=0.2 mm), which is about the same laser requirement as for polarization.

Laser cooling can be combined with a storage ring [33] or a damping ring. The current design of the 1.98-GeV NLC damping ring [34] has a circumference of 297 m, so the rotation frequency, $v_{rot}$, is ~1.01 MHz. Three NLC pulse trains of 95 microbunches each are damped for 3 interpulse periods or ~25 ms. Therefore each microbunch passes a reference point in the ring ~2.5×10$^4$ times. The proposal here is to combine laser cooling with polarization in the NLC damping ring, substituting the laser interaction for the wiggler. By combining these functions, there might be a considerable cost savings. In addition, installation of the polarization function can in principle be delayed until sometime after the damping function is commissioned.

Combined with a damping ring, a modest Nd:glass laser system plus an optical resonator as suggested by Frisch for the case of Compton scattering from an unpolarized electron beam to produce γs can be envisioned as shown in Fig. 3. For maximum polarization, the total laser energy seen by each microbunch should be ~25 J or $E_b$=1 mJ per rotation. This is a reasonable energy per optical pulse for the resonator. The average laser energy (at 1 μm) required for each microbunch in the ring is $P_{b,avg} = E_b \times v_{rot}$~1 kW. The NLC ring rf is 712 MHz with spacing between microbunches of 2.8 ns (i.e., filling every other rf bucket). There is also a gap between each pulse train. With all 285 microbunches in the ring, the laser must operate at $v_L$=357 MHz, and thus the total average power is $P_{tot,avg}= E_b \times v_L$ =357 kW. The average power required of the injection laser is $P_{inj,avg}$=100 W ($\lambda$=1 μm) operating at 357 MHz, which is probably doable. Likewise the resonator gain of $G=P_{tot,avg}/P_{inj,avg}$=3.6×10$^3$ can probably be achieved. The principal uncertainty is that the stability of an optical resonator operating under these conditions is unknown. Also the ring design will have to accommodate 2 spin rotators, and the bunch length will have to be reduced, at least in the laser interaction region, to the order of 0.2 mm. Given these complications, a new ring design, optimized for both laser damping and polarization, should be considered.

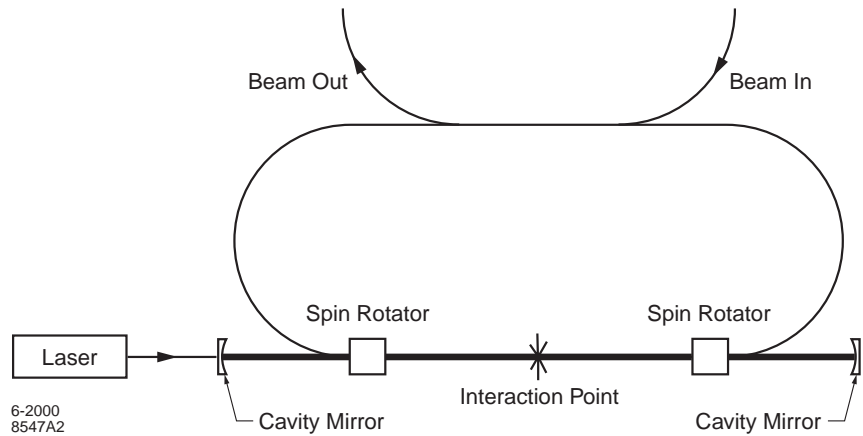

**FIGURE 3.** Conceptual layout of an NLC positron damping ring combining laser cooling and laser polarization.

# ACKNOWLEDGMENTS

The author would like to thank J. Frisch, A. Kulikov, and D. Schultz (SLAC) for numerous useful conversations with respect to positron sources and also E. Bessenov (Lebedev) and A. Potylitsin (Tomsk) for sharing their insights into the possibilities and limitations for direct polarization by Compton scattering.

# REFERENCES

1. Alley, R. et al., *Nucl. Instrum. and Meth. A* **365**, 1-27 (1995).
2. Breidenbach, M. et al., *Nucl. Instrum. and Meth. A* **350**, 1 (1994).
3. Gavrilov, N.G. et al., *Nucl Instrum. and Meth. A* **331**, ABS17 (1993).
4. Papadakis, N.H. et al., "Polarized Electrons at NIKHEF," in *Polarized Beams and Polarized Gas Targets*, edited by H.P. gen. Schieck and L. Sydow, World Scientific, Singapore, 1996, p. 323.
5. Nakanishi, T. et al., "Polarized Electron Source Development in Japan," in *Spin96 Proceedings*, edited by C.W. de Jager et al., World Scientific, Singapore, 1997, p. 712.
6. Aulenbacher, K. et al., *Nucl. Instrum. and Meth. A* **391**, 498 (1997).
7. Gowin, M. et al., "A 50kV Inverted Polarized Gun," *in Proceedings of Low Energy Polarized Electron Workshop (LE98)*, edited by Y.A. Mamaev et al., SPES-Lab-Publishing, St. Petersburg, Russia, 1998, p. 115.
8. Clendenin, J.E., "RF Photoinjectors," in *Proceedings of the XVII International Linear Accelerator Conference*, edited by C. Hill and M. Vretenar, CERN, Geneva, CH, 1996, p. 298.
9. Cornacchia, M., "The LCLS X-Ray FEL at SLAC," in *Free-Electron Laser Challenges II*, edited by H.E. Bennett and D.H. Dowell, SPIE **3614**, Bellingham,WA, 1999, p. 109.
10. Babzien, M. et al., *Phys. Rev. E* **57**, 6093 (1998).
11. Gierman, S., *Streak Camera Enhanced Quadrupole Scan Technique for Characterizing the Temporal Dependence of the Trace Space Distribution of a Photoinjector Electron Beam*, a Ph.D dissertation, University of California, San Diego, 1999, ch. 6.
12. Ferrario, M. et al., *Part. Acc.* **52**, 1 (1996).

13 Ferrario, M. et al., "HOMDYN Study for the LCLS RF Photo-Injector," contributed to the *2nd ICFA Advanced Accelerator Workshop on The Physics of High Brightness Beams*, Los Angeles, November 9-12, 1999.
14 Ferrario, M., INFN-LNF, private communication, 2000.
15 Srinivasan-Rao, T. and Smedley, J., "Table Top, Pulsed, Relativistic Electron Gun with GV/m Gradient," and F. Villa, "Acceleration of Kiloampere Current at 2.65 GV/m," in *Advanced Accelerator Concepts Seventh Workshop*, edited by S. Chattopadhyay et al., AIP Conference Proceedings 398, New York, 1997, pp. 730 and 739 respectively.
16 Srinivasan-Rao, T. et al, "Simulation, Generation, and Characterization of High Brightness Electron Source at 1 GV/m Gradient," in *Proceedings of the 1999 Particle Accelerator Conference*, edited by A. Luccio and W. MacKay, IEEE Operations Center, Piscataway, NJ, 1999, p. 75.
17 Bohn, C.L. et al., "Performance of the Accelerator Driver of Jefferson Laboratory's Free-Electron Laser," in *Proceedings of the 1999 Particle Accelerator Conference*, edited by A. Luccio and W. MacKay, IEEE Operations Center, Piscataway, NJ, 1999, p. 2450.
18 Togawa, K. et al., *Nucl. Instrum. and Meth. A* **414**, 431 (1998).
19 Nation, J.A. et al., *Proc. of the IEEE* **87**, 865 (1999).
20 Len, L.K. and F.M. Mako, "Self-Bunching Electron Guns," in *Proceedings of the 1999 Particle Accelerator Conference*, edited by A. Luccio and W. MacKay, IEEE Operations Center, Piscataway, NJ, 1999, p. 70.
21 Subashiev, A.V. and Clendenin, J.E., "Polarized Electron Beams with P≥90%, Will It Be Possible?," Preprint SLAC-PUB-8312, 2000; and Clendenin, J.E., *Int. J. Mod. Phys. A* **13**, 2507 (1998). See also Omori, T., "A Polarized Positron Beam for Linear Colliders," KEK Preprint 98-237, 1999.
22 Mikhailichenko, A.A., "Use of Undulators at High Energy to Produce Polarized Positrons and Electrons," *in Proceedings of the Workshop on New Kinds of Positron Sources for Linear Colliders*, edited by J. Clendenin and R. Nixon, SLAC-R-502, Stanford, 1997, p. 229.
23 "Conceptual Design of a 500 GeV e+e- Linear Collider with Integrated X-ray Laser Facility," edited by R. Brinkmann et al., DESY 1997-048/ECFA 1997-182.
24 Frisch, J., SLAC, private communication, 2000.
25 Okugi, T. et al., *Jpn. J. Appl. Phys.* **35**, 3677 (1996).
26 Frisch, J., "Design Considerations for a Compton Backscattering Positron Source," *in Proceedings of the Workshop on New Kinds of Positron Sources for Linear Colliders*, edited by J. Clendenin and R. Nixon, SLAC-R-502, Stanford, 1997, p. 125.
27 Dobashi, K. et al., *Nucl. Instrum. and Meth. A* **437**, 169 (1999).
28 Bessonov, E.G. and Mikhailichenko, A.A., "A Method of Polarized Positron Beam Production," in *Proceedings of the 5th European Particle Accelerator Conference*, edited by S. Myers et al., IoP Publishing, Bristol, UK, 1996, p. 1516; Potylitsin, A.P., *Nucl. Instrum. and Meth. A* **398**, 395 (1997).
29 Potylitsyn, A., "Single-Pass Laser Polarization of Ultrarelativistic Positrons," Preprint arXiv:physics/0001004, 2000.
30 Polarization of electrons in a storage ring using circularly polarized photons was proposed in Yu. A. Bashmakov, E.G. Bessonov, and Ya. A. Vazdik, *Pis'ma Zh. Tekh. Fiz.* **1**, 520 (1975), English translation *Sov. Tech. Phys. Lett* **1**, 239 (1975). See also comments on this proposal in Ya. S. Derbenev, A.M. Kondratenko and E.L. Saldin, *Nucl. Instrum. and Meth.* **165**, 201 (1979).
31 Processes not accounted for in reference 29 may affect the final polarization value. A. Potylitsyn, Tomsk, private communication, 2000.
32 Telnov, V., *Phys. Rev. Lett.* **78**, 4757 (1997).
33 Huang, Z. and Ruth, R., "Radiation Damping and Quantum Excitation in a Focusing-Dominated Storage Ring," in *Quantum Aspects of Beam Dynamics*, edited by P. Chen, World Scientific, Singapore, 1999, p. 34.
34 Corlett, J.N. et al., "The Next Linear Collider Damping Ring Complex," in *Proceedings of the 1999 Particle Accelerator Conference*, edited by A. Luccio and W. MacKay, IEEE Operations Center, Piscataway, NJ, 1999, p. 3429.